\documentclass[a4paper,12pt]{article}
\begin{document}
\vspace*{-1cm}
\hfill{\bf IReS 99-07}
\vspace{1cm}

\begin{center}

{\Large\bf Near-Threshold Production of \mbox{\boldmath $\omega$} Mesons\\[1ex]}
{\Large\bf in the \mbox{\boldmath $p\,p \to p\,p\,\omega$} Reaction}\\[5ex]

F.~Hibou,~$^1$ O.~Bing,~$^1$ M.~Boivin,~$^2$ P.~Courtat,~$^3$
R.~Gacougnolle,~$^3$ Y.~Le~Bornec,~$^3$ J.M.~Martin,~$^3$ 
F.~Plouin,~$^{2,4}$ B.~Tatischeff,~$^3$ C.~Wilkin,~$^5$
N.~Willis,~$^3$ R.~Wurzinger~$^{2,3}$\\[5ex]

$^1$ Institut de Recherches Subatomiques, IN2P3--CNRS / Universit\'e
Louis Pasteur, B.P.28, F--67037 Strasbourg Cedex~2, France\\[1ex]
$^2$ Laboratoire National Saturne, F--91191 Gif-sur-Yvette Cedex, France\\[1ex]
$^3$ Institut de Physique Nucl\'eaire, IN2P3--CNRS / Universit\'e Paris-Sud,
F--91406 Orsay Cedex, France\\[1ex]
$^4$ LPNHE, Ecole Polytechnique, F-91128 Palaiseau, France\\[1ex]
$^5$ University College London, London WC1E 6BT, United Kingdom\\[5ex]
              
\end{center}
\date{\today}

\begin{abstract}
The total cross section for $\omega$ production in the $pp\to pp\omega$
reaction has been measured at five c.m.\ excess energies from 3.8 to 30~MeV.
The energy dependence is easily understood in terms of a strong proton-proton
final state interaction combined with a smearing over the width of the state.
The ratio of near-threshold $\phi$ and $\omega$ production is consistent with
the predictions of a one-pion-exchange model and the degree of violation of the
OZI rule is similar to that found in the $\pi^-p\to n\omega/\phi$ reactions.
\\[2ex]
\end{abstract}


\newpage
\baselineskip 4ex 
 
For ideal mixing, where the $\phi$-meson is composed purely of strange quarks
and the $\omega$ contains only up and down ones, the production of the $\phi$
by non-strange hadrons is strongly suppressed by the Okubo-Zweig-Iizuka (OZI)
rule~\cite{OZI}, which forbids diagrams with disconnected quark lines.
Deviations from ideal mixing are small and these suggest that, under similar
kinematic conditions, the ratio $R$ of single $\phi$ to $\omega$ production
should be about $R\approx 4\times 10^{-3}$~\cite{Ell}. Values larger than this
could be signals for hidden strangeness in some of the incident particles or to
intrinsic violations of the rule through second-order processes. The most
striking enhancements have been seen in $\bar{p}p$ annihilation where, in
certain channels, observed $\phi$ production rates~\cite{pbp} are up to two
orders of magnitude higher than predicted by the rule.

Most reported violations are much more modest. Typical of these is the recent
comparison of $\omega$ and $\phi$ production in proton-proton collisions by
the DISTO collaboration at 2.85~GeV~\cite{DIS}. The ratio of the measured
production cross sections is
\mbox{$\sigma_{\mbox{\tiny T}}(pp\to pp\phi)/$}
\mbox{$\sigma_{\mbox{\tiny T}}(pp\to pp\omega)$}$=(3.7\pm 1.3)\times 10^{-3}$
which, after correcting for phase space effects, leads to an OZI enhancement of
about an order of magnitude. Due to the large $\phi/\omega$ mass difference,
such corrections are model dependent since the excess energy $Q$ (the kinetic
energy in the final state) is 320~MeV for the $\omega$, but only 82~MeV for the
$\phi$. While $s$ and $p$ final waves might suffice for the $\phi$, many
partial waves are likely to contribute to $\omega$ production. Any OZI test
really requires a dynamical model, but the simplest empirical approach might be
to compare rates at the same value of $Q$, where the angular momentum barriers
are similar, rather than at the same beam energy. We here report measurements
of the $\omega$ production cross section in the range $3.8\leq Q\leq 30$~MeV.
Though somewhat below the DISTO $Q$-value, these data suggest that the OZI
enhancement is little different to that seen in 
$\pi^-p\to n\omega/\phi$~\cite{Bin}.

The data were taken at the Laboratoire National Saturne as part of a program of
measuring near-threshold meson production in $pp\to ppX$ by detecting two
protons in SPESIII, a large acceptance magnetic spectrometer, and identifying
the meson $X$ as a missing mass peak. Experimental conditions were similar to
those appertaining to the production of the $\eta$ and $\eta'$~\cite{Bgd} and
so only essential features are reported.

The $\omega$ production has been measured at nominal incident proton energies
$T_p$ of 1905, 1920, 1935, 1950 and 1980~MeV using a liquid hydrogen target of
270~mg/cm$^2$ thickness. The beam intensity was measured and controlled during
the runs through an ionisation chamber placed in the beam downstream of the
target and two scintillator telescopes viewing the target. These monitors 
were calibrated using the standard carbon activation technique~\cite{Ban}.
In order to estimate the background under the expected $\omega$ peak, data
were also taken at 1865~MeV, where $\omega$ production should be negligible. 

Under standard SPESIII conditions, the momentum range of the analysed
particles is $600\leq p \leq 1400$~MeV/c, the momentum resolution
$(0.5-1.0)\times 10^{-3}$, and the effective solid angle acceptance per
particle $\Delta\Omega \approx 10^{-2}$~sr. The associated detection system,
which provides good resolution over the full acceptance of the spectrometer
together with multi-particle detection possibilities, includes three multiwire
drift chambers and four planes of scintillators as the trigger. The first of
these chambers is located near the focal surface, inclined with respect to the
optical axis of the spectrometer. Since the $\omega$ width is $8.4$~MeV/c$^2$,
particle tracks could be reconstructed, to well within the required momentum
and missing mass resolution, using information purely from this chamber.

For near-threshold meson production, it is important to verify the values of
$T_p$ or $Q$ by independent means~\cite{Bgd,Plo,Wil}. In one approach, we used
the $pp\to d\pi^+$ and $pp\to p\pi^+n$ reactions to carry out simultaneously
very accurate calibrations of the mean field of the spectrometer and
measurements of the incident proton energy at each nominal energy. The neutron
mass in the three-body final state is almost equally sensitive to $T_p$ and 
the mean field, whereas the peak position of the deuteron is much more
sensitive to the field. The experimental data were compared to the results of
extensive numerical simulations, taking into account all the details of the
experimental set-up and data-reduction algorithms. Incident proton energies 
measured by this method were slightly lower than nominal ones calculated using
the parameters of the Saturne machine, with a mean difference of $\Delta T
= T_{\mbox{\tiny nominal}} - T_{\mbox{\tiny measured}} =1.1\pm0.8$~MeV.
This offset is consistent with previous measurements~\cite{Bgd,Plo,Wil} and
we adopt it as a standard energy shift.

Typical missing mass spectra of the $pp\to ppX$ reaction are shown in Fig.~1.
That in Fig.~1a, recorded at an incident energy $T$=1865~MeV, shows a rather
uniform multipion production shape, except for the enhancement near the
kinematic limit. This arises from the combined effects of the increase in 
acceptance and the proton-proton final state interaction (FSI). Particle
production in the target windows contributes typically 10\% of the spectrum.
These data have been used to extrapolate the background underneath the $\omega$
peak to the higher incident energies using the following procedure. Let $\beta$
and $\beta_n$ be c.m.\ velocities at energies $T$ and $T_n$ respectively. The
measured momenta and angles of the protons are first transformed,
event-by-event, from the laboratory to the c.m.\ system with the velocity
$-\beta$ and then transformed back to the laboratory with the velocity
$+\beta_n$. Data recorded at $T_p$ = 2400 MeV, just below the threshold for the
$p\,p \to p\,p\,\eta'$ reaction~\cite{Bgd}, could be used to assess the
reliability of the method; the results of this test are shown in Fig.~1b. The
measured and predicted spectra are normalised to the same number of incident
protons. Since the multipion production cross section will have changed
somewhat over such a large incident energy difference, the agreement in
magnitude and especially shape are noteworthy.
 
Experimental spectra (points with error bars) obtained at $T_p$ = 1920 and
1980~MeV are presented in Figs.~1c and 1d, together with extrapolated 
background spectra (solid line histograms) and simulated $\omega$ peaks (smooth
curves). To evaluate the number of $pp\to pp\omega$ events, the extrapolated
background and $\omega$ peak were combined to fit the data in the $\omega$ peak
region and the results represented by the dashed line histogram. It should
however be noted that, at the higher energies, the background extrapolation
method does not reproduce completely the observed end-peak. The higher solid
line histogram in Fig.~1d is obtained by subtracting the simulated $\omega$
peak from the experimental spectrum. The difference between the two solid line
histograms thus represents the unexpected enhancement of the  background. At
this, the highest incident energy, the mass of the $\omega$ was used as a free
parameter to generate new background spectra. These were fit by polynomials and
an example, of degree 3, is illustrated by the smooth curve in Fig.~1d. Such
fits lead to a mid-target value of $Q= 29.1\pm1.4$~MeV which is fixed by the
production data themselves. Using this, together with $T_p = 1978.9$~MeV
determined by the pion production data, a value $m_{\omega} = 784.0\pm1.4$~MeV
is obtained. Imposing instead the compilation mass $m_{\omega}= 781.94 \pm
0.14$~MeV~\cite{PDG} gives a value of the excess energy $Q = 31.1 \pm 0.4$~MeV.
This error bar might be underestimated since, due to dynamical effects, the
apparent $\omega$ mass in hadronic production can easily change by some small
fraction of the width~\cite{Wur}. The two $Q$ determinations are broadly
consistent and their mean is quoted in Table 1 at the highest $T_p$. Since
small changes in $T_p$ could be controlled to about 0.1~MeV, the values of $Q$
closer to threshold were determined in terms of differences from this highest
point.

Very near threshold, the angular and momentum acceptances of SPESIII cover the
whole phase space for meson production in a single setting at
$\theta_{\mbox{\tiny lab}} \approx 0^{\circ}$, but angular cuts become
important for $Q > 10$~MeV. In the $pp\to pp\omega$ reaction at $T_p =
1980$~MeV, for example, only 25\% of detected events have an $\omega$ angle
lying within $45^{\circ} < \theta_{\mbox{\tiny cm}} < 135^{\circ}$. At the
energies of our experiment, the angular variation induced by higher partial
waves in the final state are expected to be small. Even at $Q = 82$~MeV the 
angular distribution of $\phi$ production is fairly flat~\cite{DIS}. 
The value of the total cross section can then be determined provided that 
the final state interaction between the two emerging protons in the $^1S_0$ 
state is taken carefully into account when determining the 
acceptance~\cite{Bgd}. The dominance of this FSI in the differential 
distributions is seen in $\eta$ production at $Q=37$~MeV~\cite{Cal}.

Another effect which must be included when estimating the acceptance for
$\omega$ production at low energy is the finite width of the resonance. The
excess energy $Q$ is defined with respect to the central $\omega$ mass value
and, at the lowest $Q$, more than 30\% of the Breit-Wigner distribution lies
below the production threshold. On the other hand, even at $Q=0$, the lower
half of the meson can be produced!

These two effects, the proton-proton FSI and the Breit-Wigner mass
distribution, were introduced into the SPESIII Monte Carlo simulation used in
calculating the acceptance. The total cross sections deduced at each mean value
of $Q$ are shown in Table~1 and Fig.~2. The errors arise mainly from
uncertainties in the number of events in the $\omega$ peak (15-27\%), the
number of incident protons, the target thickness, the detection efficiency and
the dead time (in total 12.5\%), and the acceptance (3.5-7\%), which includes
that induced by the error on $Q$.

The large momentum transfers required for heavy meson production means that the
amplitude is primarily sensitive to the short-range behaviour in the $pp$
system. In this limit, the energy dependence of the total cross section for the
production of a stable $\omega$ meson is dominated by a three-body phase space
modified by the $pp$ FSI, and this leads to
\begin{equation}
\label{1}
\sigma_T(pp\to pp\omega) = C_{\omega}\:\left(\frac{Q/\epsilon}
{1+\sqrt{1+Q/\epsilon}}\right)^{\!\!2}
\end{equation}
where, including Coulomb distortion, $\epsilon\approx 0.45$~MeV~\cite{GFW}.
Details of the production dynamics are contained in $C_{\omega}$, which is
expected to be slowly varying.

Comparing this with our data in Fig.~2, it is seen that the predicted $Q$
dependence is much too sharp. However, after smearing over an $\omega$
Breit-Wigner shape, this yields a much smoother energy dependence which
reproduces well our results with the value of $C_{\omega}= (37\pm 8)$~nb.

Direct comparison with the DISTO $\phi$ production data~\cite{DIS} is
complicated because this group has not yet deduced absolute cross sections but
only a $\phi/\omega$ production ratio. Normalising to old $\omega$ bubble
chamber data taken at slightly higher energies~\cite{HBC}, one finds that
$\sigma_T(pp\to pp\phi) = (0.28 \pm 0.14)\ \mu$b at $Q=82$~MeV. Though there
are uncertainties in extrapolating Eq.~(1) to such high $Q$, the DISTO
measurement would correspond to $C_{\phi}= (1.8\pm 0.9)$~nb. Hence
\begin{equation}
\label{2}
R_{pp}=\frac{C_{\phi}}{C_{\omega}}= (4.9\pm 2.6)\times 10^{-2}\:,
\end{equation}
to be compared with the OZI prediction of $4\times 10^{-3}$. Most of the error
in the enhancement factor of $12\pm 7$ comes from the uncertainty in the DISTO
normalisation~\cite{DIS}.

The ratio of $\phi$ to $\omega$ production has been measured near threshold
in $\pi^-p$ collisions~\cite{Bin} and, at the same value of $Q$ gives
\begin{equation}
\label{3}
R_{\pi^-p}=\frac{\sigma_T(\pi^-p\to \phi n)}{\sigma_T(\pi^-p\to \omega n)}
= (3.7\pm 0.8)\times 10^{-2}\:.
\end{equation}
An apparent $\omega$ threshold suppression, which might be kinematic in 
origin~\cite{Han}, has been corrected for.

A simple one-pion-exchange model describes {\it qualitatively} ratios of meson
production in proton-proton collisions near threshold~\cite{Bgd,CW}. In this
approach the $\phi/\omega$ ratio can be obtained in terms of that measured in
$\pi^-p$.
\begin{eqnarray}
\nonumber
R_{pp}&\approx& \left(\frac{(2m_p+m_{\omega})(m_p+m_{\phi})}
{(2m_p+m_{\phi})(m_p+m_{\omega})}\right)^{\!\!3/2}
\left(\frac{m_{\omega}}{m_{\phi}}\right)\:
R_{\pi^-p}\\
\label{4}
&\approx& 0.82\times R_{\pi^-p} = (3.0\pm 0.7)\times 10^{-2}\:,
\end{eqnarray}
where $m_p$, $m_{\omega}$ and $m_{\phi}$ are, respectively, the masses of the 
proton, $\omega$ and $\phi$.

The striking agreement between Eqs.~(2) and (4) must be considered to be rather
fortuitous in view of the large experimental {\it and} theoretical
uncertainties. Nevertheless, both the $\pi^-p$ and $pp$ experiments do suggest
strongly that the OZI enhancement is about an order of magnitude, in line with
the original DISTO analysis at the same beam energy~\cite{DIS}. Further
experimental work is clearly needed in order to establish good absolute
normalisations, but it would also help if the $\omega$ and $\phi$ data points
were measured closer in $Q$. However, understanding the theoretical
significance in the proton-proton case requires a more sophisticated model than
naive one-pion-exchange~\cite{Nak}.

We wish to thank the Saturne accelerator crew and support staff for providing
us with working conditions which led to the present results.

\newpage

%
\newpage
\noindent
\input epsf
\begin{figure}
\begin{center}
\mbox{\epsfxsize=5.3in \epsfbox{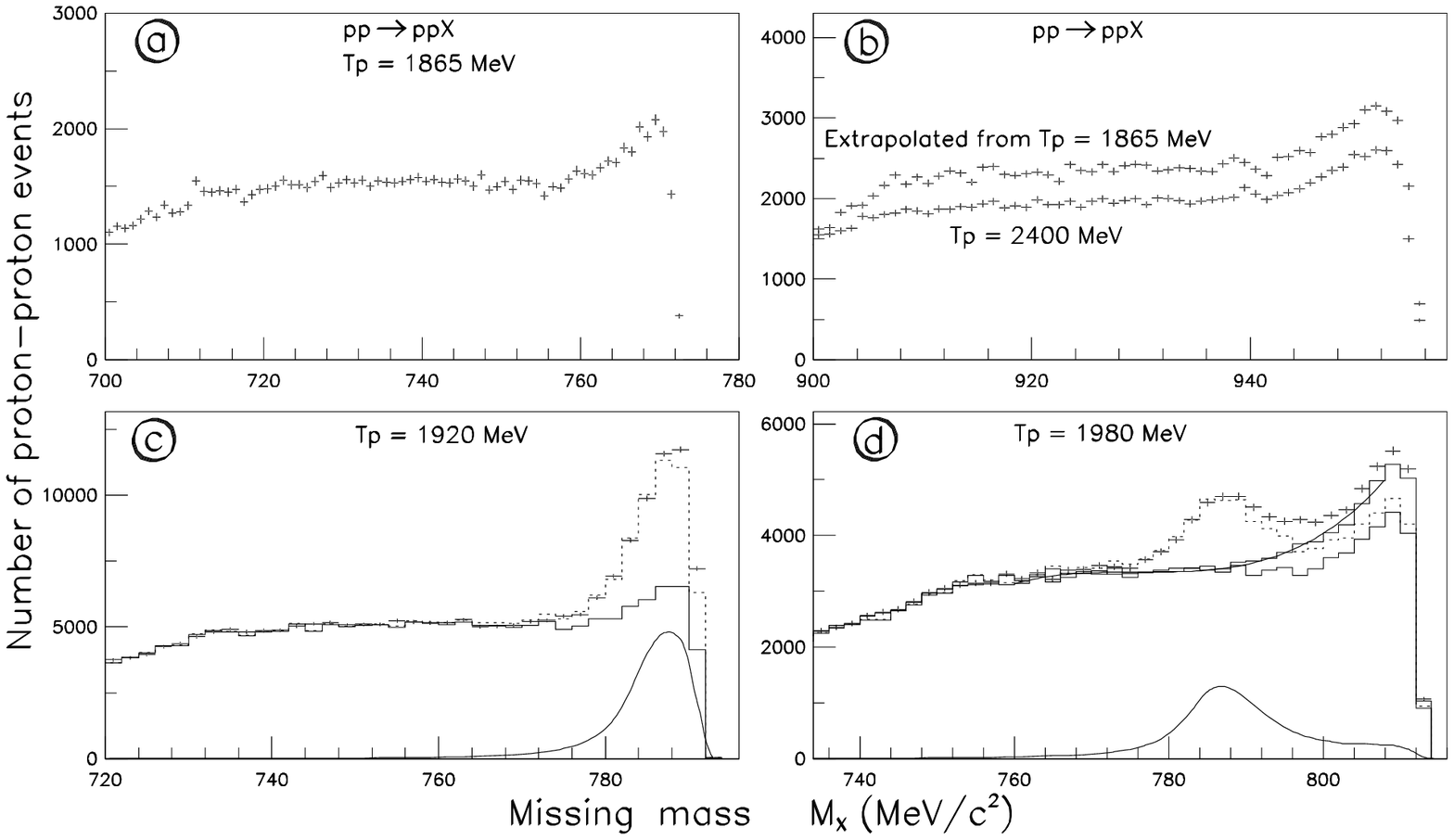}}
\vspace{5mm}

\caption{
 Missing mass spectra of the $pp\to ppX$ reaction at nominal beam
energies of (a)~1865~MeV, (b)~2400~MeV, (c)~1920~MeV and (d)~1980~MeV. The
spectrum in (a) has been used to generate the background at the higher
energies. In (b) the spectrum measured at 2400~MeV is compared to the
background extrapolated from 1865~MeV and normalised to the integrated beam
intensity. In (c) and (d) the $pp\to ppX$ spectra are shown together with
simulated $\omega$ distributions (smooth curves) and the extrapolated 
backgrounds (solid line histograms) which are combined (dashed line 
histograms) to fit the measured spectra. The higher solid line histogram 
in (d) is obtained by subtracting the simulated $\omega$ peak from the 
experimental spectrum. The smooth curve is a fit of this new background 
spectrum with a polynomial of degree 3.}
\end{center}
\label{fig1}
\end{figure}

\begin{figure}[t]
\mbox{\epsfxsize=5in \epsffile{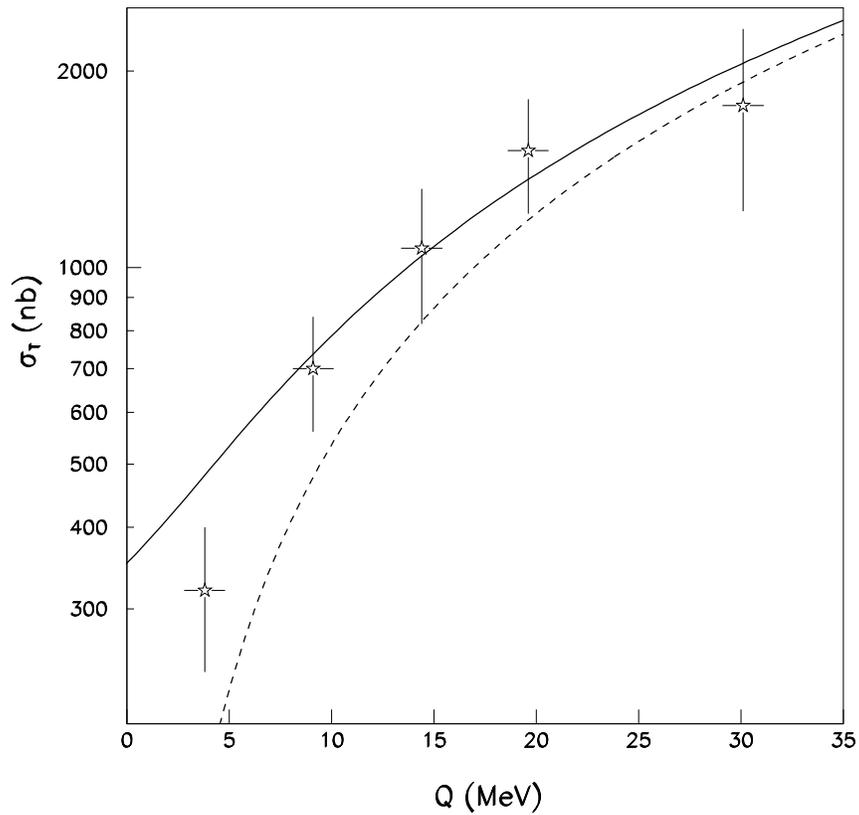}}
\caption{
Total cross section for $pp\to pp\omega$ as a function of the excess energy $Q$
with respect to the nominal $\omega$ mass. The broken curve is the prediction
of Eq.~(1) for a stable $\omega$ with $C_{\omega}=37$~nb. The solid curve
smears this result over a Breit-Wigner distribution with $\Gamma_{\omega} =
8.4$~MeV/c$^{2}$.}
\end{figure}

\begin{table}[t]
\centering
\label{table1}
\caption{Total cross sections for the $pp\to pp\omega$ reaction measured near
threshold. The excess energy $Q$ with respect to the $\omega$ central mass
value, having taken into account average energy losses of the beam in the
target, has an uncertainty of $\pm 0.9$~MeV. The quoted cross section errors
comprise all the statistical and known systematic uncertainties, including that
induced by the error on $Q$.}
\vspace{1cm}

\begin{tabular}{|c|c|}
\hline
&\\
\ \ \ \ $Q$~(MeV)\ \ \ \ \ &$\ \ \ \ \ \ \ \sigma_{\mbox{\tiny T}}~(\mu$b)
\ \ \ \ \ \ \ \\
&\\
\hline
&\\
$\:3.8$&$0.32\pm 0.08$\\
$\:9.1$&$0.70\pm 0.14$\\
$14.4$&$1.07\pm 0.25$\\
$19.6$&$1.51\pm 0.30$\\
$30.1$&$1.77\pm 0.55$\\
&\\                                                           
\hline
\end{tabular}
\end{table}

\begin{thebibliography}{99}
\baselineskip 3ex
%
\bibitem{OZI} G.~Zweig, CERN Report 8419/Th 412, 1964;
              S.~Okubo, Phys.\ Lett.\ B {\bf 5}, 165 (1965);
              I.~Iizuka, Prog.\ Theor.\ Phys.\ Suppl.\ {\bf 37-38}, 21 (1966).
%
\bibitem{Ell} H.J.~Lipkin, Phys.\ Lett.\ B {\bf 60}, 371 (1976);
              J.~Ellis, Phys.\ Lett.\ B {\bf 353}, 319 (1995).
%
\bibitem{pbp} C.~Amsler {\it et al.}, Phys.\ Lett.\ B {\bf 346}, 363 (1995).
%
\bibitem{DIS} F.~Balestra {\it et al.}, Phys.\ Rev.\ Lett.\ {\bf 81},
              4572 (1998).
%
\bibitem{Bin} D.~M.~Binnie {\it et al.}, Phys.\ Rev.\ D {\bf 8}, 2789 (1973);
              J.~Keyne {\it et al.}, Phys.\ Rev.\ D {\bf 14}, 28 (1976).
%
\bibitem{Bgd} A.M.~Bergdolt {\it et al.}, Phys.\ Rev.\ D {\bf 48}, R2969 (1993);
              A.~Taleb, PhD thesis, Universit\'e Louis Pasteur, Strasbourg
              (1994) (CRN 94-61);
              F.~Hibou {\it et al.}, Phys.\ Lett.\ B {\bf 438}, 41 (1998)
%
\bibitem{Ban} J.~Banaigs {\it et al.}, Nucl.\ Inst.\ Meth.\
              {\bf 95} 1479 (1971).
%
\bibitem{Plo} F.Plouin {\it et al.}, Phys.\ Lett.\ B {\bf 276}, 526 (1992). 
%
\bibitem{Wil} N.Willis {\it et al.}, Phys.\ Lett.\ B {\bf 406}, 14 (1997).
%
\bibitem{PDG} C.~Caso {\it et al.}, EPJ\ C {\bf 3} 633 (1998).
%
\bibitem{Wur} R.~Wurzinger {\it et al.}, Phys.\ Rev.\ C {\bf 51}, R443 (1995).
%
\bibitem{Cal} H.~Cal\'en {\it et al.}, TSL/ISV-98-0198,
              (Submitted to Phys.\ Lett.\ B).
%
\bibitem{GFW} G.~F\"aldt and C.~Wilkin, Phys.\ Lett.\ B {\bf 382}, 209 (1996).
%
\bibitem{HBC} L.~Bodini {\it et al.}, Nuovo Cimento A {\bf 58}, 475 (1968).
%
\bibitem{Han} C.~Hanhart and A.~Kudryavtsev, nucl-th/9812022
              (submitted for publication).
%
\bibitem{CW}  C.~Wilkin, in {\it Proceedings of the 8th International
              Conference on the Structure of Baryons},
              Bonn, 1998, edited by D.W.~Menze and B.~Metsch 
              (World Scientific, Singapore) p505.
%
\bibitem{Nak} K.~Nakayama {\it et al.}, Phys.\ Rev. C {\bf 57}, 1580 (1998).
\end{thebibliography}
\end{document}